\documentclass[twocolumn,secnumarabic,amssymb, nobibnotes, aps, prd]{revtex4}
\usepackage{graphicx}

\begin{document}
\title{Multiple Current States of Two Phase-Coupled Superconducting Rings}
\author{V. L. Gurtovoi$^{a}$, A. A. Burlakov$^{a}$, A. V. Nikulov$^{a}$, V. A. Tulin$^{a}$, A. A. Firsova$^{a}$, V. N. Antonov$^{b}$, R. Davis$^{b}$, and S. Pelling$^{b}$}
\affiliation{$^{a}$Institute of Microelectronics Technology and High Purity Materials, Russian Academy of Sciences, 142432 Chernogolovka, Moscow District, RUSSIA} 
\affiliation{$^{b}$Physics Department, Royal Holloway University of London, Egham, Surrey TW20 QEX, United Kingdom}
\begin{abstract} The states of two phase-coupled superconducting rings have been investigated. Multiple current states have been revealed in the dependence of the critical current on the magnetic field. The performed calculations of the critical currents and energy states in a magnetic field have made it possible to interpret the experiment as the measurement of energy states into which the system comes with different probabilities because of the equilibrium and non-equilibrium noises upon the transition from the resistive state to the superconducting state during the measurement of the critical current.
 \end{abstract}

\maketitle

\narrowtext

\section{INTRODUCTION}
A superconducting ring is a fundamental quantum object similar to an artificial atom due to the quantization of the angular momentum in the contour of the ring of all the Cooper pairs that reside in the same state. It should be noted that, unlike the atom, the superconducting ring is a macroscopic object. From the Bohr's quantization conditions for the angular momentum of the particle $mvr = n\hbar $, it follows that the difference in energies between the permitted states
$$\Delta E _{n+1,n} = E _{n+1} - E _{n} = \frac{mv _{n+1}^{2}}{2} - \frac{mv _{n}^{2}}{2} \approx \frac{\hbar ^{2}}{2mr^{2}}     \eqno{(1)}$$
decreases with an increase in the radius $r$ of the orbital or the ring (here, $v$ is the velocity). For the radius $r \approx 1 \ \mu m$ and the mass $m$ equal to the electron mass, the difference in energies $\Delta E _{n+1,n}$ corresponds to the temperature $T \approx 0.001 \ K$. Therefore, in non-superconducting rings (with $r \approx 1 \ \mu m$) fabricated from a semiconductor \cite{1PCsem} or a normal metal \cite{2PCmet}, the persistent current $I _{p}$, i.e. a mesoscopic phenomenon caused by the Bohr quantization, is observed only at the very low temperatures. In superconducting rings, this phenomenon is observed both in the rings with a large radius and at considerably higher temperatures that reach the critical temperature $T _{c}$ of this superconductor. This is associated with the fact that the quantum number $n$ of each of the Cooper pairs cannot be changed individually, and, therefore, the difference in energies (1) is multiplied by the number of pairs $N _{s}$ in the ring \cite{3AIP08}. Since this number $N _{s}(T) = Vn _{s} = 2\pi rsn _{s}(T) = 10^{5}-10^{6}$ (where $s$ is the cross-sectional area of the ring) is very large for typical sizes of the rings and $T < T _{c}$, the difference in the energies between the levels
$$ E _{n+1} - E _{n}  \approx N _{s}\frac{\hbar ^{2}}{2mr^{2}} = n _{s} \frac{\pi s}{r} \frac{\hbar ^{2}}{m}    \eqno{(2)}$$
will be of the order of $100 \ K$, which is considerably higher that the thermal energy $k_{B}T$. Therefore, with the very high probability $P _{n} \propto exp(-E _{n}/k _{B}T)$, the system occurs in the state with the lowest energy for all temperatures below $T _{c}$. The persistent current in a typical superconducting ring can reach several tens of microamperes at a temperature of $T  \leq  0.95T _{c}$. Moreover, the oscillations of the persistent current in a magnetic field are observed even in the fluctuation region at a temperature $T  \geq T _{c}$ \cite{4Science07,5Letter07}. 

With a rare exception \cite{6LT34}, in measurements of the magnetic dependence of the critical current of superconducting rings or systems of rings in the case of a weak screening, when $s \ll \lambda ^{2}$ and $LI _{p} \ll \Phi _{0} $, only one state is observed \cite{7Shift07,8JETP07}. Here $\Phi _{0} = \pi \hbar /e \approx 20.7 \ Oe \ \mu m^{2}$ is the flux quantum, $\lambda $ is the penetration depth of magnetic field, and $L$ is the inductance of the ring. Nonetheless, in the superconducting ring with a notch \cite{6LT34} or in the structure presented in Fig. 1, the energy difference can be not as great as in the case of a single ring, which enables one to investigate the energy levels nearest to the ground state. This work is dedicated to the investigation of the energy states of a two-ring structure in a magnetic field by means of the measurement of its critical current.

\begin{figure}[b]
\includegraphics{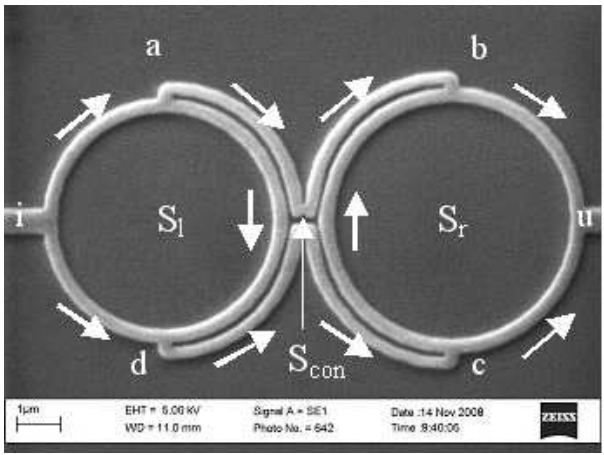}
\caption{\label{fig:epsart} Aluminium structure under investigation. The diameters of the rings are $2r _{l} \approx 4.2 \ \mu m$ and $2r _{r} \approx 4.6 \ \mu m$. The ring areas $S _{l} = \pi r _{l}^{2} \approx 13.9 \ \mu m^{2} $ and $S _{r} = \pi r _{r}^{2} \approx 16.6 \ \mu m^{2} $ correspond to the different periods of the oscillations in magnetic field  $B _{0,l} = \Phi _{0}/S _{l} \approx 1.5 \ Oe$ and $B _{0,r} = \Phi _{0}/S _{r} \approx 1.2 \ Oe$. The width of the rings is $w _{r} \approx 0.4 \ \mu m$, the width of the lines connecting the rings is $w \approx 0.3 \ \mu m$, and the film thickness is $d \approx 30 \ nm$. Arrows indicate the directions taken to be positive for the electric current and velocities of the pairs. In the regions where the external current flows, the direction from left to right is taken to be positive. In halves of the rings where the external current does not flow, the clockwise direction is taken to be positive}
\end{figure}

\section{SAMPLES AND EXPERIMENTAL TECHNIQUE}  
The structure under investigation (Fig. 1) consisted of two rings with different diameters of $r _{l} \approx 4.2 \ \mu m$ and $ r _{r} \approx 4.6 \ \mu m$ and conductors a-b and c-d connecting the phases of the rings and forming the contour with the area $S _{con} \ll  S _{l} < S _{r}$. In their central parts, the conductors had a weak links which specified the critical current of the structure. The width of the rings was $w _{r} \approx 0.4 \ \mu m$, and the width of the conductors a-b and c-d connecting the rings was $w  \approx 0.3 \ \mu m$.

The structures were fabricated by the lift-off method by depositing a thin aluminum film of thickness $d \approx 30 \ nm$ on a Si substrate using electron - beam lithography. The lithography was performed using a JEOL-840A scanning electron microscope, which was transformed into a laboratory electron lithograph by the NANOMAKER program package.  

All measurements were carried out using the four-contact method with two cryostats: (1) a glass helium (He4) cryostat, where the pumping of helium vapors made it possible to decrease the temperature down to 1.17 K; and (2) a He3 cryostat with a limiting temperature of 0.3 K. The dependence of the critical current of the structure $I _{c}$ on the magnetic field was measured from the current-voltage characteristics periodically repeated with a frequency of $5-10 \ Hz$ in a slowly varying magnetic field ($\sim 0.01 \ Hz $). A typical current-voltage characteristic is presented in Fig. 2. At temperatures below $0.99T _{c}$, the current-voltage characteristic exhibits a hysteresis; i.e., the critical currents $I _{c+}(B)$, $I _{c-}(B)$ at which the structure is switched from the superconducting state into the normal state differ significantly from the one at which it returns into the superconducting state. It should be noted that the current $I _{c+}(B)$ is not equal to $I _{c-}(B)$ because of the persistent current and an asymmetry of the structure. The transition to the resistive state occurs through an abrupt jump, Fig.2, which makes it possible to measure the critical currents with a high accuracy. Below, we present the algorithm for determining the critical currents. Three channels of a 24-bit analogue -to - digital converter with a frequency of $100 \ kHz$ simultaneously measure the magnetic field, the voltage, and the current through the structure. First, the condition is checked that the voltage $|V|$ is less than $ \varepsilon  $ (Fig. 2). This corresponds to the fact that the structure is in the superconducting state. The value of $ \varepsilon  $ exceeds the noises by a factor from 2 to 5 in order to eliminate false measurements of the critical current. When the condition $|V| > \epsilon $, which corresponds to an abrupt transition to the resistive state, is satisfied, the values of the magnetic field and the critical current measured at this instant by using other channels is written in the file. Then, the procedure is repeated. Thus, we sequentially measured the critical currents $I _{c+}(B)$ and $I _{c-}(B)$ from the current-voltage characteristics with the opposite direction of the measuring current. One dependence $I _{c+}(B)$ and $I _{c-}(B)$ containing 1000 points was measured for approximately 100 s.

In order to reduce the Earth's magnetic field, the cryostat in the region containing the sample was screened by a permalloy cylinder. The residual magnetic field was equal to $\approx  0.15 \ Oe$, and the position of the "zero" of the magnetic field was known with an accuracy of $0.03 \ Oe$ or approximately $0.02\Phi _{0}$ for the rings with a large diameter. It is well known that, in order to maintain nanostructures in a thermodynamic equilibrium at low temperatures, it is necessary to use a special electrical filtration of the wires \cite{9,10} in the case where there is a galvanic connection between the sample and the measuring system located at a higher temperature. In this work, in order to reduce noises and to thermalize samples at low temperatures, we used a filtration system consisting of low-temperature $\pi $-filters and coaxial resistive twisted pairs with the attenuation at high frequencies. Owing to the filtration, the frequency of the transmission band of each conductor leading to the sample was equal to $30 \ kHz$, and the suppression ranging from 70 to 80 dB was achieved at frequencies in the range from $100 \ MHz$ to $10 \ GHz$. Nonetheless, because of the galvanic connection of the sample at a temperature $T \approx 1 \ K$ and the measuring circuits at $T \approx 300 \ K$, both the thermal flow and electromagnetic radiation (disturbing the electronic subsystem of the nanostructures from equilibrium) were directed to the sample through the wires. The deviation of the electronic subsystem from equilibrium because of the non-equilibrium electromagnetic environment can be identified with a non-equilibrium thermal noise with an effective temperature exceeding the equilibrium temperature of the sample \cite{9}.

\begin{figure}
\includegraphics{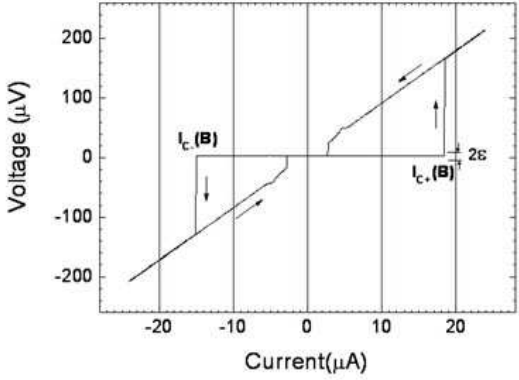}
\caption{\label{fig:epsart} Typical current-voltage characteristic used for measuring the dependencies $I _{c+}(B)$, $I _{c-}(B)$.}
\end{figure}

\section{RESULTS OF MEASUREMENTS AND DISCUSSION}
The measured magnetic dependencies of the critical current $I _{c+}(B)$ and $I _{c-}(B)$ are multi-valued and aperiodic because of three independent contours of quantization with different areas $S _{con} \ll S _{l} < S _{r}$ (Fig. 3). The single, double, and triple states of the critical current for the same magnetic field indicate permitted states $n$ with different energies $E _{n}$. Despite the fact that the structure was built as a symmetric one, we observe oscillations of the rectified voltage in the magnetic field, which are characteristic of asymmetric structures, Fig. 3. Furthermore, the maximum of the critical current are observed not at $B = 0$, but they are shifted to the opposite directions by a value $\Delta B \approx 0.25 \ G$, because of the asymmetry of the structure. The shift $\Delta \Phi \approx 0.25 \Phi _{0}$ in the extremes of the critical current is undoubtedly observed in asymmetric ring \cite{7Shift07,8JETP07}. 

This paradoxical shift $\Delta \Phi \approx 0.25 \Phi _{0}$ has no theoretical explanation. According to the fundamental requirement of the velocity quantization (see formula (1) in \cite{8JETP07}) the extremums of the oscillations $I _{c+}(B)$, $I _{c-}(B)$ can be observed only at $BS = n\Phi _{0}$ and $BS = (n+0.5)\Phi _{0}$ in the limit of the weak screening $LI _{p} \ll \Phi _{0} $ valid for the structures used both in \cite{7Shift07,8JETP07} and in our work. In particular any theoretical dependence should have the extremum at $B = 0$. We can compare our experimental results only with such dependence, Fig.4. 

The theoretical dependence $I _{c+}(B)$ shown on Fig.4 was calculated in the following way.  The transition to the resistive state occurs at $I _{c+}(B)$ when the velocity of superconducting pairs reaches the depairing velocity of $v _{c} = \hbar /m \surd 3 \xi (T)$ in one of the weak links of the conductors a-b or c-d \cite{8JETP07}. $\xi (T)$ is the correlation length of the superconductor at the temperature $T$ of measurement. The velocities on these conductors, $v _{ab} = I _{ab}/s2en _{s}$ and $v _{dc} = I _{dc}/s2en _{s}$  are determined by both the external current $I _{B} = I _{ab} + I _{dc}$ and the quantization condition of the velocity (see formula (1) in \cite{8JETP07}) in the a-b-c-d contour with a small area $S _{con}$:
$$v _{ab}l _{ab} - v _{bc}l _{bc} - v _{dc}l _{dc} - v _{ad}l _{ad} = \frac{2\pi \hbar}{m}(n _{con} - \frac{\Phi _{con}}{\Phi _{0}})$$
At the same time, the velocities $v _{ab}$ and $v _{cd}$ can be calculated from the conditions of quantization in the rings
$$v _{ia}l _{ia} + v _{ad}l _{ad} - v _{id}l _{id} = \frac{2\pi \hbar}{m}(n _{l} - \frac{\Phi _{l}}{\Phi _{0}})$$
$$v _{bu}l _{bu} - v _{uc}l _{uc} + v _{cb}l _{cb} = \frac{2\pi \hbar}{m}(n _{r} - \frac{\Phi _{r}}{\Phi _{0}})$$
and the current balance (the Kirchhoff law) for nodes a, b, c, and d:
$$ I _{ia} - I _{ab} - I _{ad} = 0, \ I _{cb} + I _{ab} - I _{bi} = 0,$$
$$ I _{dc} -I _{cb} - I _{cu} = 0, \ I _{id} + I _{ad} - I _{dc} = 0. $$

\begin{figure}
\includegraphics{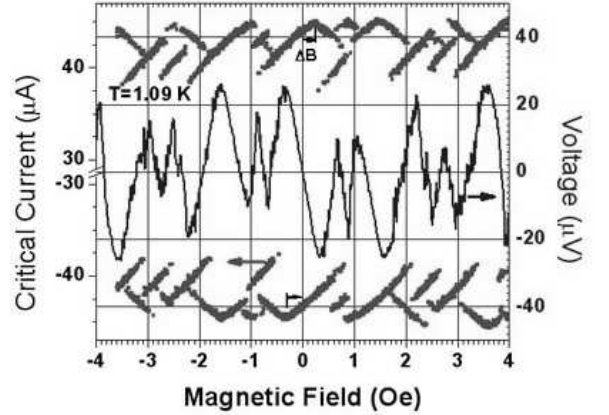}
\caption{\label{fig:epsart} Magnetic dependencies of the critical current $I _{c+}(B)$, $I _{c-}(B)$ measured at $T \approx 1.09 \ K$ and the rectified voltage arising upon application of a sinusoidal current with the amplitude $ \approx 42.5 \ \mu A$ close to the critical current of the structure at the temperature $T \approx 1.09 \ K$.}
\end{figure}

The above equations allow one to uniquely determine the velocities
$$ v _{ab} = \frac{I _{B}}{2s2en _{s}} + v _{q}  \eqno{(3a)}$$ 
$$ v _{dc} = \frac{I _{B}}{2s2en _{s}} - v _{q}  \eqno{(3b)}$$
for the given values of the external current $ I _{B}$; the quantum numbers $n _{l}$, $n _{con}$, and $n _{r}$; the magnetic fluxes inside the contours $\Phi _{l} = BS _{l}$, $\Phi _{con} = BS _{con}$, and $\Phi _{r} = BS _{r}$; and an identical density of pairs $n _{s}$ in all elements of the structure. Here,
$$ v _{q} =  \frac{2\pi \hbar}{m}\frac{(n _{l} + 2n _{con}+ n _{r}) - \frac{B(S _{l} + 2S _{con}+ S _{r})}{\Phi _{0}}}{2(l _{ab} + l _{dc}) + (s/s _{r})\pi (r _{l} + r _{r})}  \eqno{(4)}$$
is the circular velocity in the a-b-c-d contour, which is determined only by the quantization conditions for the external current $ I _{B} = 0$. The differences in the cross sections of the rings and the connecting conductors $s/s _{r} = wd/w _{r}d = w/w _{r} \approx 0.3/0.4 = 0.75$ are taken into account. When the critical velocity $v _{c}$  is reached in the weak link where the external current and the persistent current are summarised the structure as a whole jumps into the resistive state, Fig. 2. The general condition that the external current reaches the critical value is given by the expression
$$ I _{B} = I _{c} = s2en _{s}( v _{c} - |v _{q}|)  \eqno{(5)}$$
The theoretical dependencies of the critical current on the magnetic field were calculated according to formulas (4) and (5) for the sets of quantum numbers ($n _{l}$, $n _{con}$, and $n _{r}$), which minimize the energy in the given magnetic field and which were determined by a simple enumeration of all numbers (from 0 to 5) with the calculation of the energy for each possible set.

\begin{figure}
\includegraphics{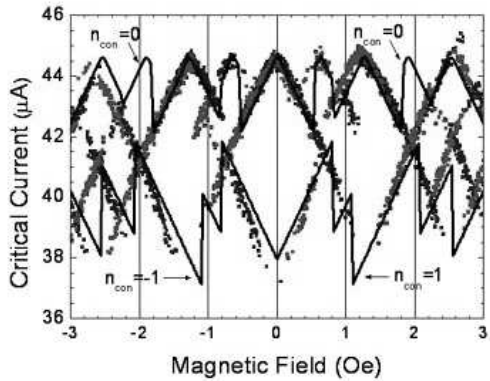}
\caption{\label{fig:epsart} The comparison of the experimental and theoretical dependencies of the critical current $I _{c+}(B)$. Points are the results of the measurements at $T \approx 1.09 \ K$, and solid lines correspond to theoretical calculations made for the system shown on Fig.1.}
\end{figure}

The kinetic energy as a function of the magnetic field was calculated by summing the contributions over all segments of the structure for the velocities that are uniquely determined by the values of the quantum numbers $n _{l}$, $n _{con}$, and $n _{r}$ and the magnetic field.

A comparison of theoretical and experimental dependencies of the critical current on the magnetic field is shown in Fig. 4. To the states with the maximum critical current (the minimum energy), there corresponds the quantum number $n _{con} = 0$, whereas the quantum number $n _{con} = \pm 1$ ($n _{con} = 1$  for $B > 0$ and $n _{con} = - 1$  for $B < 0$) determines the state with a smaller critical current (a higher energy). Despite the fact that we could not accurately reproduce some of the details of the experiment because of the complexity of the structure, the general regularities are fairly well described by the theory.

We should note one more feature in the behavior of the critical current. Figure 3 illustrates discontinuities of the critical current, which have never been observed before for both the symmetric and asymmetric rings \cite{7Shift07,8JETP07}. Apparently, these discontinuities are caused by the change in the quantum number in the rings with variations in the magnetic field. Therefore, the structures consisting of two phase-coupled rings can be used as detectors of quantum eigenstates, in contrast to single asymmetric rings, where the possibility of self-detecting these states has been predicted theoretically but not observed experimentally \cite{7Shift07,8JETP07}.

Figure 5 presents the results of the calculations for the two lower energy states of the structure and the difference in the energies $\Delta E$ between them as a function of the magnetic field. The difference in the energies of states $\Delta E$  is not very large and varies in the range from $\Delta E/k _{B}  \approx 3 \ K$   to $\Delta E/k _{B}  \approx 40 \ K$. The observation of different states in Fig. 4 at $T \approx 1 \ K$ is explained by the fact that the transition from the resistive state (Fig. 2), in which the quantum energy levels are absent, to the superconducting state, where these energy states are formed, occurs in the presence of equilibrium and non-equilibrium noises. In view of this circumstance, the system with a different probability "chooses" one of the two nearest states, which, subsequently, is detected during repeated measurements of the critical current from current-voltage characteristics. For $B = 0$, the difference in the energies is $\Delta E/k _{B}  \approx 30 \ K$; therefore, the state with a minimum energy (Fig. 4) is predominantly detected. An increase in the magnetic field up to $B =0.6 \ Oe $ is accompanied by a decrease in the energy difference down to $\Delta E/k _{B}  \approx 3 \ K$, and, as a consequence, it becomes possible to detect the system in the states with a higher energy. The probability of choosing one of the states is characterised by the density of points corresponding to the different values of the critical current (Fig. 4).

\begin{figure}
\includegraphics{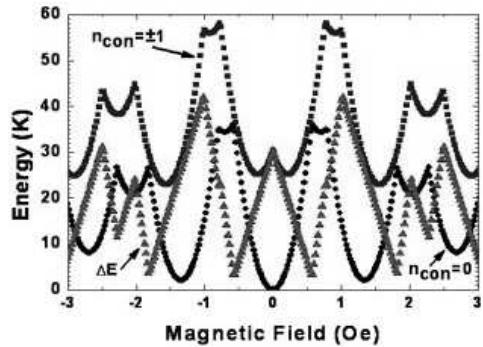}
\caption{\label{fig:epsart} The magnetic dependencies of the energies of the two nearest states corresponding to different quantum numbers of the contour with small area $S _{con}$ ($n = 0$ and $n = 1$ for $B > 0$ and $n = 0$ and $n = -1$ for $B < 0$) and the difference of these energies $\Delta E$ calculated for the experimental conditions given in Fig. 3.}
\end{figure}

\section{CONCLUSIONS}
Thus, we fabricated structures of two phase-coupled superconducting rings and investigated their energy states in magnetic fields. Multiple current states were revealed in the dependence of the critical current on the magnetic field. These states were identified with the closely spaced energy levels. The performed calculations of the critical currents and energy states in a magnetic field allowed us to interpret the experiment as the measurement of energy states into which the system comes with different probabilities because of the equilibrium and non-equilibrium noises upon the transition from the resistive state to the superconducting state. Discontinuities of the critical current were observed for the first time in ring structures without Josephson junctions, which makes it possible to use this type of structures as a detector of quantum states.

\section*{Acknowledgement}
This study was supported by the Russian Foundation for Basic Research (project no. 08-02-99042-r-ofi) and the Department of Information Technologies and Computer Systems (OITVS) of the Russian Academy of Sciences (project "Quantum Bit Based on Micro- and Nanostructures with Metallic Conductivity").

\end{document}